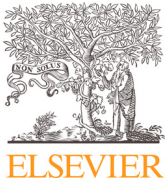

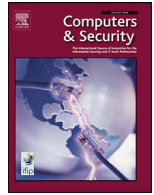

# Cryptographic ransomware encryption detection: Survey

Kenan Begovic*, Abdulaziz Al-Ali, Qutaibah Malluhi

*College of Engineering, Qatar University, Doha, Qatar*

## ARTICLE INFO

## ABSTRACT

The ransomware threat has loomed over our digital life since 1989. Criminals use this type of cyber attack to lock or encrypt victims' data, often coercing them to pay exorbitant amounts in ransom. The damage ransomware causes ranges from monetary losses paid for ransom at best to endangering human lives. Cryptographic ransomware, where attackers encrypt the victim's data, stands as the predominant ransomware variant. The primary characteristics of these attacks have remained the same since the first ransomware attack. For this reason, we consider this a key factor differentiating ransomware from other cyber attacks, making it vital in tackling the threat of cryptographic ransomware. This paper proposes a cyber kill chain that describes the modern crypto-ransomware attack. The survey focuses on the Encryption phase as described in our proposed cyber kill chain and its detection techniques. We identify three main methods used in detecting encryption-related activities by ransomware, namely API and System calls, I/O monitoring, and file system activities monitoring. Machine learning (ML) is a tool used in all three identified methodologies, and some of the issues within the ML domain related to this survey are also covered as part of their respective methodologies. The survey of selected proposals is conducted through the prism of those three methodologies, showcasing the importance of detecting ransomware during pre-encryption and encryption activities and the windows of opportunity to do so. We also examine commercial crypto-ransomware protection and detection offerings and show the gap between academic research and commercial applications.

## 1. Introduction

The incursion of digital and online lifestyles in almost every segment of our lives has brought multiple consequences related to dependence on integrity and availability of information in business and personal matters. One of those consequences is our inability to live, work or even receive life-dependent services like medical treatment or water and electricity supply if related digital resources and data are unavailable or compromised. Cybercrimes are seeing significant growth across all geographies, with ransomware being the leading type of attack (Singleton et al., 2021). Ransomware is a type of attack where malicious actors utilize multiple tactics and techniques to gain the capability to lock or encrypt a victim's data. This attack usually results in an ultimatum where the victim-user either pays for unlocking or decryption keys or faces losing all their data. Due to the already mentioned dependency on digital lifestyle, data is constantly growing in importance, creating an environment for a very lucrative business for ransomware gangs since the first recorded attack in 1989. While crypto-ransomware is a more common type of attack and lock-ransomware is in the decay (Berrueta et al., 2019), the latter is still relevant, especially in the mobile platforms (Su et al., 2018). According to a Fortinet survey, ransomware grew by 1070% across different industry verticals between July 2020 and June 2021 (Fortinet, 2021). Critical services like the health sector, especially in the age of the COVID-19 pandemic, have been particularly vulnerable and targeted—the U.S. Health and Human Services Department has tracked 82 ransomware attacks in the first five months of 2021. The average cost of the incident in the U.S. health sector was around USD1.27 million, even though only USD131,000 was the average cost of the ransomware payment (U.S. Department of Health and Human Services Cybersecurity Program, 2021). The rest of the cost was distributed across lost business costs, including increased customer turnover, lost revenue due to system downtime, and the increasing cost of acquiring new business due to diminished reputation. Depending on the industry and geography, in other sectors worldwide, the ransom ranged between USD7.75 million and USD0.37 million, making the average cost of ransomware incidents in 2021 USD1.85 million (Sophos, 2021). The ransomware threat goes even further, with the Conti ransomware group announcing their support for the Russian invasion of Ukraine at the end of February 2022 and ac-tive participation in cyber warfare utilizing their capabilities and the available access to various assets worldwide (Russia-based ransomware group Conti issues warning to Kremlin foes | Reuters).

* Corresponding author.
  *E-mail address:* kb2000115@qu.edu.qa (K. Begovic).

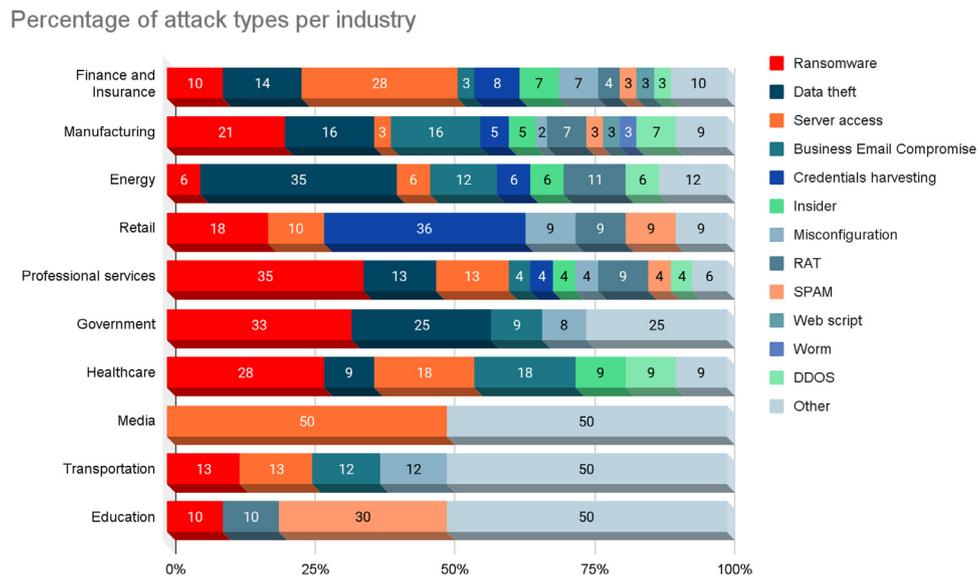

**Fig. 1.** Percentage of attack types per industry (Singleton et al., 2021).

Despite all the reporting and high-profile cases of ransomware attacks, they continue to flourish and grow in sophistication and effectiveness. The reason for this probably lies in the fact that, according to Fortinet's survey in 2021, 96% of companies that were already victims of ransomware gangs responded that they were moderately ready for the ransomware attack, even though 16% of them suffered from three or more attacks (Fortinet, 2021).

As shown in Fig. 1, in industries like Professional Services, Government, and Healthcare, the percentage of ransomware attacks as a portion of all cyber attacks is 35%, 33%, and 28%, respectively, making this type of attack by far the most common attack overall (Singleton et al., 2021).

Nevertheless, another trend was noticed in Sophos' research on the state of ransomware. The malicious ransomware actors are moving away from generic and automated large-scale attacks to more targeted attacks executed with precision and persistence (Sophos, 2021). A review of the available data on modes of ransomware groups' operation points to apparent similarities with the Advanced Persistent Threat modus operandi. This observation partially explains the increase in the difficulty of detecting and defending against these attacks compared to defending against malware like common viruses, trojans, or worms.

In the targeted crypto-ransomware attack, the malicious actor uses various techniques to gain the capability to encrypt the victim's data. Such techniques evolve, becoming more focused (Sophos, 2021) and using precise no-noise attacks on the networks (Wang et al., 2018). Despite the shifting of techniques and some tactics, cryptographic ransomware carries one differentiating characteristic that separates it from malware: the capability and goal of encrypting victims' data so that only malicious actors can decrypt it upon the ransom payment.

In this survey, existing proposals of pre-encryption and encryption detection techniques were reviewed to show their importance in countering ransomware and the possibility of being the ultimate solution for eliminating this threat. Detecting and countering crypto-ransomware has long been at the forefront of scholarly research. With the advent of the COVID-19 pandemic, motivation for ransomware attacks increased, and research interest in this topic has grown to an ever-larger extent. Most pre-encryption and en-cryption detection solutions operate in a host-based environment focusing on file system and kernel activity monitoring. However, some detection solutions focus on network communication inside local target networks and communication with command and control servers. The latter algorithms do not necessarily utilize network information to detect DNS-based indicators of compromise (IOC) but also deep packet inspection to detect cryptographic key delivery and exfiltration. The comprehensive set of algorithms and techniques to detect pre-encryption and encryption varies from simple decoys placement and file integrity monitoring to complex machine learning (ML) models trained on monitoring systems' behavior during encryption and encryption-related operations, such as key generation. The survey also focuses on encryption-related detection in crypto-ransomware, and any further references to ransomware are related to the encryption of victims' data by malicious actors with the purpose of extortion.

After introducing the topic of cryptographic ransomware, this paper covers related survey-like works available at the time of writing in the section 1.1 *Related Work*. Further, we propose a cyber kill chain to describe cryptographic ransomware attacks and discuss each of the defined phases in the kill chain, describing the behavior and methodology of attacks. In the survey part of the paper, we review research on the detection of activities related to the Encryption phase as described in the discussion of the proposed cyber kill chain. We also provide a brief survey of commercial solutions and usage of encryption detection outside of the crypto-ransomware use case.

### 1.1. Related work

Several surveys related to ransomware have been published, primarily focusing on defining the characteristics of ransomware attacks. However, there were no previous attempts to build a survey of detection techniques related to encryption as a hallmark of ransomware attacks.

Recent literature on ransomware threats is largely focused on three main streams. The first stream revolves around identifying recent ransomware threats based on static and dynamic analysis developed by the scientific community. The second stream aims to classify ransomware threats without necessarily focusing on detection algorithms. Finally, the third stream engages with holistic ap-proaches to ransomware techniques and tactics. The following will briefly present these studies.

With regard to the first stream, Moussaileb et al. (2021), in their survey of ransomware threats to Windows operating systems, have unified all detection techniques based on static and dynamic analysis developed by the scientific community since 2014. This survey treats both crypto and locking ransomware types and, despite the title and general topic of the paper, covers some Android ransomware cases as well. The existing surveys focus on crypto ransomware strictly (Berrueta et al., 2019), noting the difficulty of surveying this novel topic since data from various papers is impossible to compare due to different metrics and approaches to ransomware.

Regarding studies focusing on the second stream, in an earlier attempt to survey research on ransomware, Al-rimy et al. (2018) provided a comprehensive classification of ransomware attacks but with few details on detection algorithms. Also worth mentioning is a paper by Eze et al. (2018), that attempted a holistic examination of ransomware techniques and tactics in a very general and brief manner. Other survey-like papers focus on the evolution of the ransomware phenomenon (Zavarsky and Lindskog, 2016) or actual empirical data about real-world attacks (Connolly et al., 2020). In their survey of ransomware detection solutions, Herrera Silva et al. (2019) focus on identifying and listing all the detection and prevention parameters identified in the surveyed research, and they consider situational awareness concerning the same.

About the third stream, which takes more innovative approaches, more comprehensive surveys (Oz et al., 2022) cover all available varieties of platforms targeted by ransomware and consider the historical context and chronology of ransomware development. Other similar works (Dargahi et al., 2019) take the systematization of ransomware features' taxonomy as a center of their proposal, and, similar to ours, the authors propose a cyber kill chain that attempts to describe and encompass all ransomware behavior observed so far. Also, some proposals focus on certain operating systems like Android (Ameer et al., 2018), Windows (Moussaileb et al., 2021; Reshmi, 2021; Naseer et al., 2020), or methods and tools in detection like a machine and deep learning and big data (Urooj et al., 2022; Bello et al., 2021). Finally, some proposals seek to build benchmarks for researchers who want to introduce more innovative approaches in ransomware detection mechanisms (Maigida et al., 2019).

Cryptographic ransomware detection has interested the academic community and the cybersecurity industry. Methodologies and techniques for detection use static and dynamic analysis of components and actions belonging to the cryptographic ransomware lifecycle phases. Some focus on local user machines, user and program activities, and the state of files in memory and file systems. Others look at the network indicators of ransomware presence, ranging from detecting single ransomware based on its signature to complex heuristic techniques and machine learning algorithms looking at multiple stages of the ransomware lifecycle.

Digging deeper into the available literature, it is noticeable that only some research papers focus on the issue of encryption in crypto-ransomware. Those usually concentrate on machine learning algorithms (Kok et al., 2020a) or methods like frequency of encryption estimation (Mülders, 2017). Furthermore, approaches focusing on the state of files in the file system (Jethva et al., 2020; Jung and Won, 2018), monitoring of the hardware performance (Dimov and Tsonev, 2020), and even the energy consumption (Azmoodeh et al., 2018) show promising results in detecting encryption.

This paper aims to survey contributions to the research of *encryption detection* in ransomware and techniques valuable for detecting the ransomware Encryption phase. The analysis does not employ first-hand information like in some other more general surveys on the crypto-ransomware (Berrueta et al., 2019). Instead, it focuses on results in other scientific and industry-based propositions with a strong focus on encryption detection. The outline of the contributions of this paper relative to the recent ransomware surveys can be summarized as follows:

- Compared to other survey papers in the field, this survey provides a deeper dive into the detection of encryption by compartmentalizing the detection of encryption techniques and treating them as independent cases, even if they are part of a hybrid solution.
- We identify a widening gap between richly-diverse academic literature on the detection of encryption techniques on the one hand and commercial implementations in market-leading solutions on the other.
- We provide an overview of some of the key challenges and, in our view, misconceptions when approaching the topic of crypto-ransomware.
- We present the need for a better organized cyber kill chain that describes the modern crypto-ransomware attack.
- We propose a needs-based, field-informed contemporary cyber kill chain.

For completeness, an apt description and classification of cryptographic ransomware attacks in their methodologies and phases will be presented with a brief classification of detection techniques.

## 2. On crypto-ransomware behavior and methodology

The crypto-ransomware attack is characterized by a specific action of encrypting victims' data with the intention to extort financial or other benefits as a ransom for decryption. Researchers have observed distinct actions that mark noticeable separate phases of a ransomware attack (Moussaileb et al., 2021; Berrueta et al., 2019; Al-rimy et al., 2018; Eze et al., 2018). After a careful examination of different proposals for ransomware-specific kill chains, as well as the growing tendency of ransomware groups to carefully choose the target and emulate Advanced Persistent Threat (Sophos, 2021), we synthesized our findings and, as a result, identified four distinct phases of a crypto-ransomware attack. Our proposal for a kill chain is shown in Fig. 2.

### 2.1. Phases of the attack

There are many other surveys of ransomware used kill chains with different numbers and scopes of phases. We propose a kill chain with four distinct steps or phases for a ransomware attack. The kill chain, presented in Fig. 2, was found to be the best fit to focus on detecting encryption as a defining characteristic of cryptographic ransomware attacks. The following section will explain in detail the essential characteristics of each of the four phases of our kill chain, namely Initial compromise, Establishing foothold, Encryption, and Extortion, to present ransomware's lifecycle and emphasize the importance of the Encryption phase.

#### 2.1.1. Initial compromise

Initial compromise marks the phase in which a ransomware attack compromises the first computer. Various methods for delivering and executing initial compromise include phishing, spearphishing, corrupt web pages, and actual security bugs and system misconfigurations (vulnerabilities). Fig. 3 shows the most common methods of initial compromise based on original research by the authors, covering the years between 2013 and 2021.

As presented in Fig. 3, phishing is the most common method for initial compromise, often combined with exploiting vulnerabilities or corrupted websites.

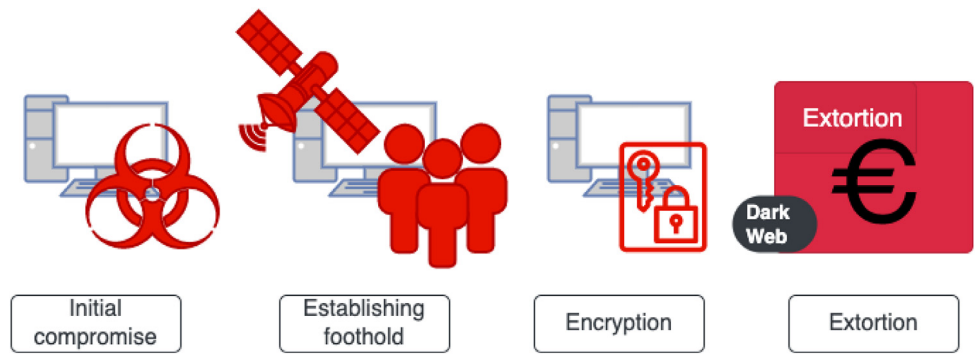

**Fig. 2.** Ransomware kill chain.

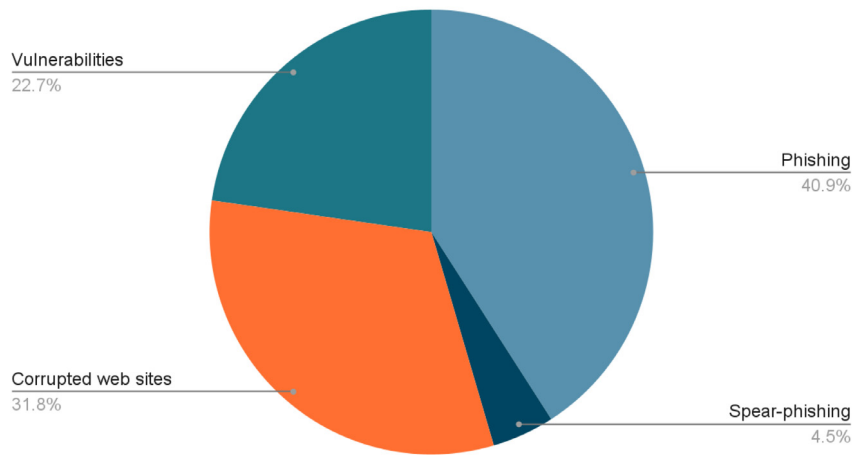

**Fig. 3.** Initial compromise attack vectors.

Spear-phishing is rare, but that could result from many researchers not focusing on segregating spam (unsolicited emails), phishing, and spear-phishing.

Due to non-standardized terminology describing the initial compromise techniques, applying the same unique valuation across all sources takes a lot of work. The result is that some ransomware attacks fall into more than one category. However, for the sake of clarity, we placed such ransomware attacks into the category corresponding to its most defining characteristic.

### 2.1.2. Establishing foothold

In most cases, after the initial compromise, the attacker attempts to establish a permanent foothold in the compromised system and move laterally or otherwise. The activity usually, but not necessarily, starts with connecting to command and control (C2) servers. C2 is an Internet host or entire infrastructure built to control ransomware's behavior, issuing commands, generating, distributing, and/or storing encryption keys, and collecting information about the ransomware victim.

Ransomware attacks that do not utilize C2 reduce detection surface to the host detection capabilities only, entirely avoiding network detection measures that focus on communication detection between the initial intrusion code and C2 (Berrueta et al., 2019). If this type of ransomware propagates and establishes a foothold in the manner of a worm, then network controls can detect it (Alotaibi and Vassilakis, 2021). Notable ransomware attacks that do not use C2 are BadRabbit (Alotaibi and Vassilakis, 2021), CT-BLocker (Upadhyaya and Jain, 2016), Bart (Labs, 2017a), KillDisk (The rise of TeleBots, 2016), Patcher (New crypto–ransomware hits macOS, 2017), Revenge (Revenge Ransomware, a CryptoMix Variant, Being Distributed by RIG Exploit Kit), Spora (Lemmou et al., 2021), BTCWare (Wood and Eze, 2020), Crysis (Wood and Eze, 2020), NotPetya (NotPetya Ransomware Attack [Technical Analysis], 2017), GlobeImposter (Dargahi et al., 2019), Sage2.0 (Sage 2.0 Ransomware), Scarab (Lemmou et al., 2021), LockerGoga (Adamov et al., 2019), Jigsaw (Berrueta et al., 2019), Ryuk (A Targeted Campaign Break-Down - Ryuk Ransomware, 2018), and Zeoticus 2.0 (Walter).

Ransomware attacks that use C2 to establish control over compromised hosts and further direct actions use three different approaches such as C2 server static IP address, static DNS domains for C2 servers, and dynamically generated domain names. With static C2 server IP addresses, the IPs to which ransomware attempts to connect in this attack phase are already hard encoded within attack tools and files downloaded in previous stages. In a recent example, the ransomware dubbed Maze was widely distributed in Italy during 2020 using the list of static IPs to connect to C2 servers and share the information about the victim host immediately after the encryption (Ransomware Maze, 2020). Unlike Maze ransomware, WannaCry used static hard-encoded DNS domains to access C2 servers instead of IP addresses. Incidentally, another static DNS domain, iuqerfsodp9ifjaposdfjhgosurijfaewrwergwea.com, was hard-coded into WannaCry. Subsequently, the ransomware researchers found the domain name to be a kill-switch for WannaCry propagation (Akbanov et al., 2019). Finally, dynamically generated domain names characterize ransomware families that aim to make both static binary analysis and network detection difficult. Examples like

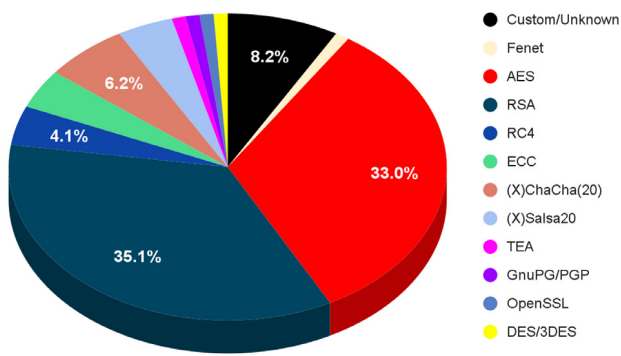

**Fig. 4.** Usage of encryption algorithms by major ransomware families 1989 - 2021

Locky and TeslaCrypt ransomware (Berrueta et al., 2019) utilize dynamic generation algorithms (DGA) to create domain names dynamically. DGA's purpose is to make it difficult for defenders to discover and block C2 servers' names and/or IP addresses. In order to keep its activity hard to detect and yet avoid total randomness, the DGA is using some of the following building elements:

- Seed, which can be a word(s) and/or number(s), is a building element introduced by ransomware DGA writers, and it can be changed to segregate C2 domain names between different versions or groups of victims.
- Time-based is the element that changes dynamically with time. It does not need to be necessarily influenced by time or date, and some other event can trigger it; the only condition is that it changes over a period of time.
- Top-level domains (TLDs) are the final part of DGA-created domain names. The first two create the body of a domain name by being combined, and then a predetermined TLD is added. TLDs like ".xyz," ".top," and ".bid" are very popular when creating DGA (Arntz, 2016).

Ransomware C2 servers' communication plays a prominent role in many proposed ransomware detection mechanisms that detect C2 IPs and domain names in the ransomware tools and network traffic. These can be used in activities from deny-listing all the way to detecting DGA-created domain names in DNS queries to be used with DNS sinkholes (Dynamic Resolution: Domain Generation Algorithms, Sub-technique T1568.002 - Enterprise | MITRE ATT&CK®).

### 2.1.3. Encryption

The Encryption phase of a ransomware attack includes the following phases: encryption key generation, obtaining a public key from the C2 server, searching file system, encryption, exfiltration of data with specific extensions or in particular folders, and deletion of possible backups like shadow volumes.

Different ransomware families use various encryption schemes to encrypt their victims' data. Whether the attacker chooses to use symmetric, asymmetric, or a combination of both encryptions directly influences cryptographic key generation and management during the Encryption phase of the attack. Table 1 names prominent ransomware families since 1989 and their choice of encryption. The distribution of cryptographic methods with symmetric and a combination of symmetric and asymmetric are most commonly used, while asymmetric alone is used much less. While researching sources for information contained in Table 1, the authors have compiled data from these sources to create Fig. 4, which shows the distribution of various encryption algorithms' usage from the first ransomware attack in 1989 to the end of 2021. In the case of exclusive symmetric encryption use, key generation is done by either using local operating system cryptographic capabilities or a custom implementation of cryptographic algorithms.

In Microsoft Windows, ransomware uses the function BCrypt-GenRandom Cryptography API: Next Generation (CNG), as exemplified by Noberus ransomware (Noberus) or CryptGenRandom - Maze ransomware (Ransomware Maze, 2020). In Apple's macOS and IOS, the SecRandom function carries similar capabilities to CryptGenRandom and Linux, along with several other UNIX-like operating systems that implement getrandom as a system call. Ransomware for the latter operating systems uses open source libraries like mbedtls - examples seen in

KeRanger (New OS X Ransomware KeRanger Infected Transmission BitTorrent Client Installer, 2016) and RansomEXX (RansomEXX Trojan attacks Linux systems, n.d.) ransomware re. The secret key is sometimes protected when utilizing an asymmetric encryption scheme in remote secure storage. In the case of the local generation of keypair, the secret key is encrypted with another C2-provided public key. The public key is either locally generated with a secret key, supplied by a C2 server, or both. Ransomware like Cerber used C2 supplied RSA public key to encrypt locally generated RSA secret key that was used to encrypt locally generated RC4 key used for victim's files encryption (Sala). On the other hand, CryptoWall ransomware would not start encryption unless a 2048-bit RSA key is received from C2 (Cabaj et al., 2015).

In most cases, successful ransomware attacks combine symmetric ciphers like Rijndael, ChaCha/Salsa20, or RC4 together with asymmetric ciphers like RSA or ECC. This is primarily due to the speed of encryption advantage that symmetric cipher provides over asymmetric encryption when encrypting a large volume of data. In scenarios where the secret key remains on C2, asymmetric encryption is a good option to encrypt the symmetric key. This way, the victim's responders to that attack would not be able to use it in decryption before paying the ransom. The speed is also a factor in locating the files to be encrypted by ransomware. Some attackers infect all drives alphabetically (in Windows-based attacks), while some limit infection to specific user folders like Desktop or Documents. Most sophisticated ransomware provides whitelist exclusion of specific system folders and system configuration files to maintain the operating system's functionality after the encryption (Lemmou et al., 2021).

During the actual encryption, ransomware applies four tactics: reading, encrypting in memory, writing to the file system, and removing original files. While reading a file, ransomware like CryptoWall tries to read files in one read, reducing the number of read/write operations (Lemmou et al., 2021). On the other hand, ransomware can use fixed block lengths for reading and writing files during encryption. WannaCry or LockerGoga ransomware read files in 256 kb and 64 kb blocks, respectively (Loman, 2019). The third approach to encryption is when ransomware performs a read of the fixed buffer from the beginning or from the end of the file twice before committing a write to the file system. This behavior has been observed in ransomware Spora which uses two read operations checking for ransomware added specified values to the content of each file before encryption from the end to establish if the file is already encrypted (Lemmou et al., 2021).

Finally, ransomware can write directly to the original file and then optionally rename it during the destruction of the original files. Another way of destruction is by saving encrypted files in the new location and then deleting, moving, or overwriting the original. Also, the third method includes moving the original file to some temporary location, overwriting it with encrypted data, and then moving it back to its original place in the file system.

RIPlace is a new technique of replacing the original files with encrypted files that have been able to bypass all of the known protection systems for the Windows family of operating systems (CISOMAG, 2019). Found in ransomware like Thanos (Walter), the

**Table 1**

Major ransomware families' usage of encryption schemes during the Encryption phase of the attack in chronological order 1989 - 2021.

| Encryption scheme | Ransomware families in chronological order |
| --- | --- |
| Symmetric encryption | AIDS trojan (PC Cyborg) (Case Study), GPCode (Emm, 2008), Cryzip (Cryzip Ransomware Trojan Analysis), MayArchive (MayArchive.B Description | F-Secure Labs), SympLocker (Ameer et al., 2018), TeslaCrypt (Lemmou and E. M. Souidi, 2018), Tox (Meet "Tox," 2015), TorrentLocker (Wyke and Ajjan, 2015), DMALocker (Raheem et al., 2021), Xorist (Reshmi, 2021), Jigsaw (Conti et al., 2018), Cerber (Pletinckx et al., 2018), CryptXXX (Berrueta et al., 2019), Enigma (Berrueta et al., 2019), Bart (Reshmi, 2021), Fsociety (Berrueta et al., 2019), Raa (Berrueta et al., 2019), Satan (Reshmi, 2021), Spora (Labs, 2017b), KillDisk(Linux) (Conti et al., 2018), CryptoShadow (Reshmi, 2021), DoubleLocker (Lipovský et al., 2018), CryptoShield (Berrueta et al., 2019), Patcher (Ransomware Recap), Revenge (GoldSparrow, 2017), BTCWare (Wood and Eze, 2020), Erebus(Win) (Ransomware Recap), WannaLocker (Hu et al., 2020), Gibon (GIBON Ransomware), Locker (Berrueta et al., 2019), Retwyware (Reshmi, 2021), Scarab (Berrueta et al., 2019), Netwalk (Take a "NetWalk" on the Wild Side, 2020), Try2Cry (Try2Cry Ransomware - IBM X-Force Collection), EKING (Zhang, 2020), Conti (Conti Ransomware), LV (LV Ransomware), 54BB47H/Rollcast/Arcane (Kitten.gif: Meet the Sabbath Ransomware Affiliate Program, Again | Mandiant) |
| Asymmetric encryption | Archievus (Case Study), CryptoDefense (Herzog and Balmas, 2016), CryptoWall (Cabaj et al., 2015), VirLock (crypto version) (Zavarsky and Lindskog, 2016), CryptVault (Berrueta et al., 2019), Linux.Encoder (Berrueta et al., 2019), Chimera (Conti et al., 2018), SamSam (Berrueta et al., 2019), GlobeImposter (Berrueta et al., 2019), Hermes2.1 (Shevchenko et al., 2017), Katyusha (Reshmi, 2021), Thanos (Thanos Ransomware, 2020), Hive (Walter), N3tw0rm (N3TW0RM ransomware emerges in wave of cyberattacks in Israel) |
| Combination of symmetric and asymmetric encryption | GPCode (Blackmailer), CryptoLocker (Hansberry et al., 2014), CTBLocker (Weckstén et al., 2016), DMALocker4.0 (Raheem et al., 2021), Locky (Almashhadani et al., 2019), Petya (Aidan et al., 2017), KeRanger (Conti et al., 2018), Anubis (GoldSparrow, 2016), Matrix (Threat Assessment, 2021), KillDisk(Win) (Conti et al., 2018), Sage (Labs, 2017c), Erebus(Linux) (Erebus Resurfaces as Linux Ransomware, 2017), WannaCry (Chen and Bridges, 2017), Crysis (Crysis Ransomware Gaining Foothold, Sets Sights to Take Over TeslaCrypt - Wiadomości bezpieczeństwa), NotPetya (NotPetya Ransomware Attack [Technical Analysis], 2017), BadRabbit (Alotaibi and Vassilakis, 2021), GandCrab (Usharani et al., 2021), Ryuk (Umar et al., 2021), LockerGoga (Adamov et al., 2019), PewCrypt (Jegede et al., 2022), Zeppelin (blogs.blackberry.com), PXJ (PXJ Ransomware Campaign Identified by X-Force IRIS, 2020), REvil (Analyzing the REvil Ransomware Attack, 2021), Maze (Ransomware Maze, 2020), SunCrypt (When Viruses Mutate, 2021), Cyrat (Hahn, 2021), SMAUG (SMAUG Ransomware), Mount Locker (Mount Locker Ransomware In The Mix - IBM X-Force Collection), Fonix (Walter), Exorcist (Velasco, 2020), Egregor (Egregor Ransomware, Used in a String of High-Profile Attacks, Shows Connections to QakBot, 2020), Zeoticus2.0 (Walter, n.d.), RansomEXX (RansomEXX Trojan attacks Linux systems), BlackMatter (Updated, 2021), Grief (Dark Web Threat Profile, 2021), BlackByte (BlackByte Ransomware – Pt. 1 In-depth Analysis), Karma (Karma Ransomware | An Emerging Threat With A Hint of Nemty Pedigree - SentinelOne) |

RIPlace utilizes IRP_MJ_SET_INFORMATION system callback in combination with the legacy DefineDosDevice function to delete original files. At the same time, renaming is performed on both original and encrypted files.

Deletion of backup files most commonly occurs with the deletion of Windows Volume Shadow Copy using operating system tools or through encryption of shared drives when some sort of NAS solution is deployed for backup purposes.

*2.1.4. Extortion*

Once the files are entirely or, in some cases, partially encrypted, the ransomware creates a ransom note as a text or HTML file instructing the victim on what to do in order to retrieve their data.

Payment of ransom in the extortion phase of ransomware attack has represented difficulty for cyber-criminals since ransomware's first appearance in 1989. The inability to remain anonymous has pushed early ransomware attackers to use payment means like premium-rate text messages or pre-paid vouchers like Paysafe cards (Oz et al., 2022) in the times before the appearance of cryptocurrency. After the introduction of BitCoin in 2009, most ransomware attackers moved towards cryptocurrency ransom payments in the Extortion phase of the attack. In 2012, locker ransomware Reveton was the first Ransomware-as-a-Service (RaaS) and the first ransomware to demand payment in BitCoin. Among cryptographic ransomware, CryptoLocker in 2013 was the most advanced and among the first to strongly emphasize payment by Bit-Coin (Liao et al., 2016).

The section 2.1 has outlined the main characteristics of all four kill chain phases. We identified the most common instances of crypto-ransomware behavior and methodology. However, in order to adapt this kill chain into actionable recommendations necessary for the effective prevention of ransomware, the following sections will introduce a novel approach where the focus in detecting ransomware is concentrated on the detection of Encryption as conceptualized in the previous section.

## 3. Detection of encryption

Research in detecting ransomware in general through various phases of attacks has snowballed in the past several years. Focused research on cryptographic ransomware follows a general trend of the tremendous increase in published research; however, most surveys remain focused on all attack phases described previously in section 2.1. Encryption is the defining characteristic of a crypto-ransomware attack. The usage of different encryption algorithms, as shown in Fig. 4, choice of symmetric, asymmetric, or combination (hybrid) of different encryption schemes, as shown in Table 1, shows how cryptographic ransomware closely followed the trend in its evolution and how encryption itself continues to be the one differentiating characteristic that is the most obvious candidate to be the factor in the detection of the attack.

When researching the phenomenon of cryptographic ransomware through time, we observe that despite the evolution of this threat from malware to something similar to advanced persistent threat (APT), encryption remained a uniquecharacteristic that separates this ransomware from other information security threats. The capability to encrypt without any control gives ransomware attackers the primary motivation and purpose for executing the attack. With that in mind, we surveyed and classified methodologies used to detect ransomware while operating inside the Encryption phase of the attack. The scholarly literature on ransomware detection largely clusters around the following three major groups:

- API and system call monitoring-based detection
- I/O monitoring-based detection
- file system operations monitoring-based detection that include
  - scanning for high entropy in files and
  - monitoring deception tokens in file system-based detection.

Machine learning (ML), even though sometimes covered as a separate methodology in encryption detection, is cross-cutting the three previously mentioned methodologies depending on informa-

**Table 2**

A list of the most significant Windows API calls for crypto-ransomware detection was collected from surveyed papers.

| API Calls | Description |
| --- | --- |
| FindFirstChangeNotificationA | Creates a change notification handle and sets up initial change notification filter conditions. Await on a notification handle succeeds when a change matching the filter conditions occurs in the specified directory or subtree. |
| SHEmptyRecycleBinA | Empties the Recycle Bin on the specified drive. |
| SHFileOperation | Copies, moves, renames, or deletes a file system object. |
| SHBrowseForFolder | Displays a dialog box that enables the user to select a Shell folder. |
| SHLoadInProc | Creates an instance of the specified object class from within the context of Shell's process. |
| SHGetFileInfo | Retrieves information about an object in the file system, such as a file, folder, directory, or drive root. |
| SHQueryRecycleBinA | retrieves the Recycle Bin's size and the number of items in it for a specified drive. |
| SHPathPrepareForWriteA | Checks to see if the path exists. This includes remounting mapped network drives, prompting for ejectable media to be reinserted, creating the paths, prompting for the media to be formatted, and providing the appropriate user interfaces, if necessary. |
| SetUserFileEncryptionKey | Sets the user's current key to the specified certificate. |
| EncryptFileA | Encrypts a file or directory. All data streams in a file are encrypted. All new files created in an encrypted directory are encrypted. |
| DecryptFileA | Decrypts an encrypted file or directory. |
| OpenEncryptedFileRawA | Opens an encrypted file in order to backup (export) or restore (import) the file. This group of Encrypted File System (EFS) functions is intended to implement backup and restore functionality while maintaining files in their encrypted state. |
| FileEncryptionStatusW | Retrieves the encryption status of the specified file. |

tion used as features in the dataset that the ML model is trained on. The usage of machine learning in ransomware detection is prevalent in hybrid proposals that employ data from different groups. The same is true for detecting activities in the Encryption phase of the attack. While some of the proposals are included in the survey for their specific use and discussion of machine learning issues related to the detection of ransomware pre-encryption and encryption activities, often in combination with other methods, categorization to each major group that was described previously was done based on the models' input features. ML ransomware encryption detection includes machine learning techniques and algorithms used to detect encryption and related activities by ransomware attacks in hybrid and pure implementations.

The following section presents the state-of-the-art methodologies used to detect ransomware inside the Encryption phase of the attack.

### 3.1. API and system calls monitoring

Monitoring of function and system calls to detect activity related to file encryption aims at detecting encryption at early stages. At this point, API functions and system calls related to encryption operations appear in the dynamic monitoring of events in an operating system or static analysis of binary files. Depending on the operating system, the dynamic detection mechanism focuses on API functions or system calls, and static analysis employs a more comprehensive observation for a set of function calls and text strings. This method is often part of a hybrid solution for ransomware detection. While proposing machine learning solutions for static and dynamic analysis of files suspected of being ransomware, Sheen and Yadav (2018) identified ransomware's most commonly used API calls. Table 2 presents those API calls in detecting the Encryption phase of ransomware attacks with a brief explanation of their purpose in the Windows operating system. Their solution's performance was measured on the ML model's success in differentiating between benign and malicious utilization of all defined features. In contrast, these API functions were also found in other research.

Some authors propose a detection model for ransomware using monitoring API calls and developing pre-encryption detection algorithms (Yadav et al., 2021). Even though their paper outlines methods and goals, it still needs to provide concrete solutions. Others use NLP (Natural Language Processing) using convolutional neural networks to analyze API sequences retracted from both ransomware and benign processes in the secure sandbox (Qin et al., 2020). Observing sequences of API calls in machine learning solutions was also applied by Ahmed et al. (2020), in their en-hanced Minimum Redundancy Maximum Relevance methods, and the most relevant features fine-tuned from each were not necessarily encryption-related. Similarly, Almousa et al. (2021) consolidated Windows operating system API calls found in 51 collected ransomware families with common API calls in software to train machine learning models to detect ransomware. In the case of Mehnaz et al. (2018), with their RWGuard proposal and CryptoAPI Function Hooking (CFHk) Module, they implement a technique of intercepting CryptoAPI calls by hooking function through memory address space, shifting and mandatory JMP instructions intercepts, and securely stored CryptoAPI activities. Kok et al. (2020a) focus their research on extracting API calls in various ransomware samples prior and the call of APIs containing the word 'crypto' while the sample was running in a sandbox. Their pre-encryption detection algorithm used a machine learning model trained on the extracted API calls dataset. The limitation of their algorithm is its reliance on Windows API calls which disable them in order to detect ransomware by using custom encryption functions. Another example of looking at pre-encryption API calls is the work of Al-Rimy et al. (2020), who established a two-component detection system consisting of DynamicPre-encryption Boundary Definition (DPBD) and Features Extraction (FE). The former creates the pre-encryption boundary vector with all cryptography-related APIs used to create the boundary of the pre-encryption activities that define the boundary of pre-encryption. The latter extracted all relevant pre-encryption features for use in detection. Arabo et al. (2020) observation of API calls by DLLs in Windows in their hybrid solution proposal pointed out the correlation between process behavior and ransomware activity in the pre-encryption and encryption stages. A similar proposal, focusing on the Android operating system, came from Scalas et al. (2019), who argued that using System API observation in detection systems performs better than complex solutions that combine more complex ransomware indicators.

On the static observation of API functions related to encryption, Xu et al. (2017) proposed CryptoHunt deals with obfuscation in binary files that prevents detecting Windows Cryptographic API or OpenSSL functions. While this proposal was resilient to various obfuscation techniques, some crucial elements of the process, detecting custom cryptographic functions, were not described. API monitoring and system calls are also widely used as input features in machine learning-based detection of activities in the Encryption phase. In their proposal, Al-Rimy et al. (2021) argue that feature extraction in the early Encryption phase of a ransomware attack and phases before creating a situation of too few data and high dimensional features leads to a substantial risk of overfitting. As a remedy, they proposed "a novel redundancy coefficient gradual up-weighting approach." The calculation of redundancy terms of mutual information was introduced to improve the feature selection process and enhance the accuracy of the detection model. The experiment showed better accuracy with the proposed approach by testing multiple classifiers in all cases.

Similarly, Hwang et al. (2020) propose two-stage detection of crypto-ransomware, first building the Markov model from Windows API call sequence patterns capturing the characteristics of ransomware behavior and then using a Random Forest classifier over remaining data (registry keys operations, file system operations, strings, file extensions, directory operations, and dropped file extensions) to control false-positive and false-negative rates. Their two-stage mixed detection model gives 97.28% overall accuracy, 4.83% false-positive rate, and 1.47% false-negative rate. Also, Kok et al. (2020b), as an extension to their already mentioned proposal of the Pre-encryption Detection Algorithm (PEDA), proposed a set of conventional and unconventional metrics for PEDA's learning algorithm (LA) component performance. By introducing metrics like Likelihood Ratio (LR), Diagnostic odds ratio (DOR), Youden's index (J), Number needed to diagnose (NND), Number needed to misdiagnose (NNM), and net benefit (NB), they improved the performance in this unique use case when compared to using only conventional metrics.

Al-rimy et al. (2019) focused on pre-encryption activities detection in their proposal for a crypto-ransomware detection model. They proposed two combined approaches, the first incremental bagging (iBagging) technique and enhanced semi-random subspace selection (ESRS), which act as an ensemble model. iBagging creates subsets depending on the observation of ransomware behavior, while ESRS then creates subspaces that were used to train a pool of classifiers. The best classifiers were modeled using a grid search, and a voting system was employed. While accuracy was higher than in competing approaches for the same datasets, there were limitations related to feature selection in different subspaces. Since the features were selected within one subspace independently, the same feature could be selected in more than one subspace. This decreases the accuracy of the model.

Although this literature is very developed and ever-growing, it still needs a comprehensive focus on researching the detection of Encryption phase-related activities in the case where encryption algorithms are custom-implemented using third-party libraries. The development of static analysis methods, as well as the discovery of patterns in API and system calls for dynamic analysis when custom encryption implementation is used, would significantly improve the overall success of this method.

### 3.2. I/O (input/output) monitoring

Monitoring I/O is another technique that monitors internal behavior in an operating system. It aims to use information from I/O requests related to memory, file system, and even network for the detection of ransomware encryption (and other phases). Like API and system calls monitoring, this technique is often part of a more comprehensive detection solution involving several different methodologies and techniques. Kharaz et al. (2016) used a combination of techniques and methodologies in their proposal for a detection system named UNVEIL. Their monitoring of I/O established that regular applications could generate I/O access requests generated by ransomware encryption tools. However, due to the common design where these regular applications do not block access to the original files, their sequence patterns of I/O operations differ from ransomware. The paper also describes zero-day ransomware detection by observing entropy between read and write operations. McIntosh et al. (2021), as a part of the broader proposal for an access control framework, utilized I/O monitoring us-ing various models of the framework name RANNACCO. Their solution nested its modules between Windows I/O and Storage class driver. While the overall framework proposed has limitations, the I/O monitoring part is said to detect encryption successfully. RW-Guard by Mehnaz et al. (2018) implements a hybrid solution with IRPParser that logs I/O requests and passes them further to other modules.

Network monitoring is also used to detect pre-encryption and encryption events. Almashhadani et al. (2019) proposed network-based crypto-ransomware detection using Locky ransomware as the case study. Their proposal was built using multiple independent classifiers over both packet and flow data. Using a total of 18 features that were extracted from TCP, HTTP, DNS, and NBNS traffic, the proposal achieved 97.92% accuracy for packet-based and 97.08% accuracy for flow-based data. By using TCP and UDP features computed from network flows, Fernández Maimó et al. (2019), as a continuation of their previous proposal for an integrated clinical environment named ICE++, proposed a machine learning detection and protection system that was capable of anomaly detection and ransomware classification. It also uses Network Function Virtualization (NFV) and Software-Defined Networking (SDN) paradigms to prevent the spread of crypto-ransomware activity. They trained multiple models using multiple algorithms and achieved a precision/recall of 92.32%/99.97% in anomaly detection and an accuracy of 99.99% in ransomware classification. Roy and Chen (2021) proposed a solution named DeepRan that prevents the spreading of the Encryption phase across network-connected computers. Deep-Ran utilizes attention-based bi-directional Long Short Term Memory (BiLSTM) with a fully connected layer to model the normalcy of networked hosts. Its behavior anomaly detection processes substantial amounts of logging data collected from bare metal servers. Conditional Random Fields (CRF) model was used to extend BiL-STM for detected anomalies to be classified as potential ransomware attacks. Semantic information extraction from "high dimensional host logging data" was done by the Term Frequency-Inverse Document Frequency (TF-IDF) method. Early ransomware detection had a 99.87% detection accuracy (F1-score of 99.02%).

On the hardware monitoring level, Paik et al. (2016) propose monitoring I/O for encryption detection in addition to their SSD hardware monitoring. Similarly, Dimov and Tsonev (2020) monitor HDD performance and utilize the I/O performance rate for disk read and disk write operations to detect ransomware in the Encryption phase. Finally, as a unique type of I/O monitoring, Park and Park (2020) propose hardware tracing for detecting symmetric key cryptographic routine detection in malicious binaries that employ anti-reverse engineering techniques. Azmoodeh et al. (2018) proposed detecting crypto-ransomware activity in IoT by monitoring power consumption and applying machine learning models to the collected data. Their proposal employs Dynamic Time Warping (DTW) as a distance measure with KNN as a classifier, outperforming conventional classifiers like Neural Networks, KNN, and SVM. Their approach achieved a detection rate of 95.65% and a precision rate of 89.19%.

Only a few research papers that include I/O monitoring in relation to detecting activities related to the Encryption phase of a ransomware attack are available. The success of some of the mentioned proposals in detecting encryption as activity and overall events that are a consequence of the Encryption phase indicates that much more can be done in this field and that more innovative hybrid solutions are possible.

### 3.3. File system monitoring

Monitoring file system activity to detect the Encryption phase of ransomware attacks focuses on collecting information about the state of the file system and the files themselves. The early ideas, like Young et al. (2012), to use different sector hashes to detect target files, including encrypted files, paving the way for file system usage in the fight against cryptographic ransomware.

By using raw binary files as ML features, Khammas (2020) proposed a Random Forest classifier that uses 1000 n-gram features extracted directly from raw bytes using frequent pattern mining. The selection of features was made using the Gain Ratio to reduce the dimensionality of features. The proposal maintains an optimal number of trees to be 100 with achieved accuracy of 97.74%. While this proposal focused on the analysis of binaries that contain ransomware attack tools and recognition of the same among benign binaries, due to the nature of crypto-ransomware behavior, it is safe to assume that many of the n-gram features were related to the Encryption phase. Furthermore, using APK files containing the source code of an Android app, Sharma et al. (2021) extracted features from the file related to ransomware attacks. Their proposal, named RansomDroid for detecting crypto-ransomware activity in Android devices, uses an unsupervised machine learning model. Unlike K-Means clustering, the proposal used a Gaussian Mixture Model with a flexible and probabilistic approach to modeling the dataset. Feature selection and dimensionality reduction for improvement of the model were also utilized. The model detects Android ransomware with an accuracy of 98.08% in 44 ms. Almomani et al. (2021) also use analysis of Android APK files for feature extraction in their proposal. They rely on an "evolutionary-based machine learning approach" to detect cryptographic ransomware in Android devices. They used the binary particle swarm optimization algorithm (BPSO) to tune the classifier's hyperparameters and feature selection. Synthetic minority oversampling technique (SMOTE) with support vector machine (SVM) algorithm was used for classification. The combination name SMOTE–tBPSO-SVM used g-mean as a metric and achieved a result of 97.5%.

Tang et al. (2020) proposed a detection and prevention system named RansomSpector that monitors the file system and network activities. It is a virtual machine-based system that resides in the hypervisor, thus making it difficult to bypass through privilege escalation. The crypto-ransomware was detected with extraordinarily little overhead to performance - less than 5%. Continella et al. (2016), in their proposal for ShieldFS, offer a whole new file system, which, combined with the machine learning portion of their proposal, can detect ransomware behavior as an anomaly, including operations related to the Encryption phase. Similarly, Lee et al. (2022) proposed statistical analysis to differentiate between regular and encrypted blocks in the file system. Their solution, Rcryptect, utilizes extracted heuristic rules using FUSE (File system in Userspace) to avoid kernel modification. Rcryptect, among the other methods, detects high entropy files created by cryptographic operations. Nevertheless, the solution faces common issues of false positives for benign files with high entropy and the issue that prevention mechanisms can cause damage to some files under attack before the ransomware encryption process is killed. Entropy, in combination with fuzzy hashing, as a means to detect files encrypted by ransomware in the file system was proposed by Joshi et al. (2021). They used a mini-filter driver that interacts with file system behavior as kernel mode. While achieving more than 95% of detection success in their experiment, the method is susceptible to explorer.exe process DLL injection that would bypass the security measures proposed. Lee et al. (2019) use entropy estimation to detect files encrypted by ransomware in a cloud environment. When using the cloud as a backup, there exists a risk that encrypted files could be synchronized to the cloud. The authors thus observed the number of ransomware encryption attacks and divulged the baseline used in entropy estimation over files in the cloud. Their experiment reported a 100% success rate in detecting encryption. Jung and Won (2018) used the entropy of files in their comprehensive ransomware detection and protec-tion system. They utilized context-aware analysis that used information from APIs, file system metadata, systems to detect large-scale read/write operations, and entropy analysis capable of detecting benign usage of encryption with enhanced classification to improve entropy analysis results. Similarly, Jethva et al. (2020) proposed their system to detect and prevent crypto-ransomware using entropy in multilayer detection. The technique was combined with monitoring registry key operations, file signatures in the Windows operating system, and machine learning. By improving the already mentioned method of analyzing the entropy of files, Hsu et al. (2021) examined 22 different file formats of encrypted files and extracted features to be used with the Support Vector Machine algorithm. They achieved a detection rate of 85.17% using the SVM Linear model, which increased to over 92% when using the SVM kernel trick (with the polynomial kernel) model.

Not all of the research favors entropy use in detecting ransomware encryption. McIntosh et al. (2019) propose depreciation of this method in the fight against ransomware, arguing that techniques they identified to mitigate entropy usage in encryption detection are sufficient to invalidate reliance on entropy information. In their experiment, BASE-64 encoding and partial file encryption have shown their effectiveness in "confusing" entropy information; thus, the usage of File Integrity and File Type Identification have been proposed as alternatives to using entropy measures.

RWGuard is a proposal by Mehnaz et al. (2018) that combines multiple techniques in the real-time detection of cryptographic ransomware. The solution includes monitoring the file system for malicious activity through File Monitoring and File Classification modules. Also, it has the capability to automatically generate decoy files in the file system using a feature called Decoy Files Generator. Some of the techniques used for detection are inherently probabilistic and prone to false positives. Another decoy-in-the-file-system-based solution is the proposal of R-Locker by Gómez-Hernández et al. (2018), a novel approach to creating *honeyfiles* as decoys to detect and stop ransomware in action. To achieve this in practical implementation on UNIX-like operating systems, a new named pipe is created in the file system containing a specially crafted small-size honeyfile. An alert is raised to kill the process when ransomware attempts to read the file. Due to the fact that the kernel manages synchronization, these regular applications do not block access to the original files between reading and writing; if a process attempts to read more data than expected, reading is blocked until the writer makes up for expected data. This feature allows time to raise alerts about suspected ransomware behavior. In another approach using deception, honeypots for IoT devices were proposed by Sibi Chakkaravarthy et al. (2020) based on Social Leopard Algorithm (SoLA) to model honey folders. The Intrusion Detection Honeypot (IDH) also introduces an Audit Watch module that monitors the entropy of files in the device, together with a module called Complex Event Processing (CEP) that collects information from multiple external security sources used to confit and stop ransomware activity. SoLA algorithm is critical to this proposal with its capabilities to process extracted features from processes that accessed the honey folder.

Usage of entropy to detect encryption is present in various operating systems, including Android. A proposal by Jiao et al. (2021) detects custom encryption with an accuracy rate of 98.24% in Android platforms using only entropy information.

In a more ML-focused approach, using the activity logs for features extraction that contain all of the filesystem events, Homayoun et al. (2020) proposed applying Sequential Pattern Mining to find Maximal Frequent Patterns (MFP) in logged activities for known ransomware. This created candidate features to be used in classification by multiple machine learning classification algorithms. In their experiment, authors used J48, Random Forest, Bag-ging, and Multi-Layer Perceptron (MLP) classifiers and achieved an F-measure of 0.994 with a minimum AUC value of 0.99 in the detection of ransomware samples from benign activities using Windows registry, DLL, and file system to registry log of activities. F-Measure of more than 0.98 with a false-positive rate of less than 0.007 in the detection of a given ransomware family using 13 selected features whose significance was recognized during the research. Their results were not short of impressive, creating datasets of ransomware logs for 1624 ransomware binaries sourced from virustotal.com, as well as separate sampling for overfitting. However, no indication was given that testing was performed on an independent dataset.

Even though an impressive amount of research has been found in relation to file system monitoring for the purpose of detecting the Encryption phase activities in a ransomware attack, we feel that research went truly little in the direction of tying the techniques mentioned above into access control systems used to control file systems. Entropy detection is an auspicious tool in fighting cryptographic ransomware; more proposals are needed on this topic.

### 3.4. Commercial solutions brief survey

Most of the commercial offering for protection against crypto-ransomware focuses on providing capabilities in the area of Enterprise Backup and Recovery. Companies that focus solely on ransomware protection in their products are almost nonexistent. Many vendors focus on delivering recovery capabilities through air-gapped backup and immutable backup copies, and detection is based chiefly on integrity and anomaly behavior monitoring. Even though vendors offer markets as comprehensive data protection solutions, it is indicative that most of them emphasize that the backup is the last line of defense against ransomware, which in some instances indicates that other ransomware protection controls are expected to be in place for the product to live to its expectations.

According to Gartner's Magic Quadrant for Enterprise Backup and Recovery Software Solutions (Rao et al., 2021), the major capability for the evaluation of a product was Ransomware detection and protection. An example is Acronis (Ransomware Protection with Backup for Business - Acronis) which offers both cloud-based and on-premise solutions that include the capability to actively scan for ransomware activity and verify the authenticity and recoverability of backup copies. Another major player in this field is Arcserve (Ransomware Protection Solution for an Impenetrable Business) which does not have its own capabilities to detect and protect against crypto-ransomware built into its product but rather has excellent cooperation with security giant Sophos that provides the capability for them. Cohesity (Ransomware Recovery | Reduce Downtime with Rapid Recovery) is the leader in enterprise backup and recovery. They offer cloud service with immutable backup using the write-once-read-many (WORM) feature and RBAC access control model. Also, they utilize machine learning for anomaly behavior detection to detect crypto-ransomware activity. Commvault Ransomware Recovery - Commvault offers one of the most comprehensive lists of capabilities against ransomware. Their approach employs a zero-trust model, built architecture using NIST's Cybersecurity Framework (CSF). Their detection capability mostly relies on anomaly detection in both networks and file systems. Dell Technologies (Dell EMC Cyber Recovery Solution – Cyber and Ransomware Data Recovery), another leader in this vertical, provides detection using their Intelligent CyberSense Analytics. It utilizes machine learning for anomaly detection. Other important vendors and leaders in this area, like Veeam (Ransomware Protection: Learn How Veeam Can Protect Your Data) and Rubrik (Ransomware Re-covery), use similar techniques, and there are no serious differentiating factors in detecting crypto-ransomware.

Other groups of vendors that focus on ransomware detection are traditional threat detection and response companies. They rely on anomaly detection utilizing various monitoring techniques that provide hybrid solutions of API calls and I/O monitoring, and file system changes. Some are using machine learning models, and there are occasional claims of artificial intelligence (AI) that are difficult to confirm. The most significant ones are Carbon Black (Endpoint Protection Platform | VMware Carbon Black Endpoint), Trend Micro (Enterprise Ransomware Protection & Removal), Darktrace (Darktrace for Ransomware), Extrahop (Ransomware Mitigation & Detection Solution - ExtraHop) , and Vectra AI (Ransomware Detection and Response - Ransomware Solutions | Vectra AI).

Concerning the relationship between academic research achievements surrounding the detection of ransomware executing the pre-encryption and encryption activities and commercial solutions, it has been noted and observed that an apparent discrepancy exists between the two (Scala et al., 2019; Nicol et al., 2015). The persistence of these differences is not uncommon in cybersecurity-related topics and has been driven by a series of factors categorically branched into categories. Therefore, we have factors that are technical, procedural, or bureaucratic in origin and nature. Amongst those most substantially addressed by the literature are the factors identified as distinctly technical in origin:

- Integration

    Industry solutions utilize different detection methods and measures in a coordinated schema of safeguards, where a comprehensive set of firewalls, intrusion prevention systems, and endpoint protection defenses are adopted to combat real-world threats. Antithetically, the academic approach necessitates the separation of specific detection techniques and their study in isolation, in effect, obfuscating principles of translation.

- Scalability

    Detection solutions developed in an isolated academic environment tend to need help with monitoring and analysis capabilities for the vast amounts of data generated by modern industry networks (Scala et al., 2019). Forasmuch as the complexity and diversity of commercial networks are not to be understated; scaling assumes an integral role in the application of research outcomes in the real-world market (Nicol et al., 2015).

- Complexity

    The algorithms developed through mechanisms of academic inquiry often involve complex and resource-intensive modus operandi, which fail to be practical for real-world deployment or, within the nature of their construction, cannot evolve into more flexible setups (Nicol et al., 2012).

- Adaptability

    With standardized models built on specific ransomware samples and behavior patterns in training academically developed technologies, limitations arise in adaptability for real-world applications (Scala et al., 2019). Ransomware, in real world attacks, rarely follows the uniformity established in these test models. On the contrary, its behavior constantly evolves in a race to outmaneuver newly forged preventative measures as they emerge.

Likewise, while procedural factors present similar practical limitations as those found with technical factors, the nominal core of these limitations lies in qualities inherent to the experimental process and research objectives rather than market translation:

- Validation

  The process of testing and validation notably differ across academic and industry borders. Research validation roots itself in the accuracy of simulations and the control of experimental conditions. Commercial solutions require rigorous real-world validation against various extraneous variables and incorporating existing infrastructure to ensure effectiveness, reliability, and compatibility (Grossman et al., 2001; Nicol et al., 2015).

- False positives

  Concerning research goals, the process of addressing false positives should be addressed to achieve high detection rates (Bold et al., 2022; Kok et al., 2020b). Comparatively, greater emphasis is placed on false positives in the industrial context, as the losses incurred through the resultant disruptions are of greater interest. Hence, commercial solutions seek a compromise between adequate detection accuracy and minimizing false positives.

- Time to market

  Rapid response is necessary to combat the emergence rate of new threats and meet the demands of the commercial market. Concurrently, academic research development, testing, and refinement require heavy time investments for the analytic procedure, translating to a natural lag in the pace of ransomware evolution (Kashef et al., 2023; Nicol et al., 2015).

In contrast to technical and procedural factors, which are determined by objectively measurable discrepancies, bureaucratic factors are interpretational, based on institutional inefficiencies of anthropogenic origin. The most effective of which present themselves in restrictions of:

- Intellectual property and licensing

  Hesitancy in adopting newly developed technologies based on academic research is often directly tied to the risks involved with an investment in new IPs where precedents have yet to be set on the extent of its protected status. Similar contingencies arise with licensing restrictions that introduce additional expenditures in the form of permissions and approvals.

### 3.5. Other applications of detection of encryption

Detection of encryption is not always related to cryptographic ransomware defense. There are numerous use cases where it would be necessary to recognize if encryption happened or is happening.

An example is Ameeno et al. (2019) proposal using the Naive Bayes algorithm to differentiate between compression and encryption and identify file types. Li and Liu (2020) proposed an encryption detection method using deep convolutional neural networks (CNN). The proposal uses converted raw data into two-dimensional matrices as an input to CNN. The results showed a higher detection rate than competitive storage and network encryption methods. In another proposal that utilizes the power of machine learning and deep learning, Yang et al. (2021) propose the usage of natural language processing (NLP) in combination with the two in detecting encrypted network traffic. The technique usually used for weighting in message retrieval and keyword extraction, Term Frequency-Inverse Document Frequency (TF-IDF), was used in modeling detection due to its capability not to need analysis of each field in network traffic. Both ensembles of various machine learning classifiers and CNN were used with high accuracy and their own advantages. The advantage of CNN in deep learning is that it efficiently deals with sparse matrices (compressed or uncompressed) generated by TF-IDF in situations where there is an "abundance" of hardware resources.

On the other hand, encrypted traffic detection in limited hardware resources is better suited for ensemble classifiers. Finally, even though somehow similar to crypto-ransomware encryption detection, a proposal by Dong et al. (2021) named MBTree deals with the detection of encrypted traffic between Remote Access Trojan (RAT) and Command & Control server (C2). The proposal relies on building the kind of baseline by integrating flow-level DirPiz sequences as a synthesis of host-level Multi-Level Tree (MLTree). Actual detection was done by measuring path similarity and node similarity of actual traffic with the baseline. The F-1 score reported is 94%.

## 4. Conclusion

In closing, our journey through the myriad proposals relating to the detection of pre-encryption and encryption activities by cryptographic ransomware has surfaced some findings and lessons learned. The dynamic nature of ransomware necessitates constant vigilance, innovation, and adaptation. Each of the surveyed methodologies has its own advantages and disadvantages. This all is the reason we present section 4.1, Lessons learned. Our findings in section 4.2 underscore the importance of multi-layered, robust detection mechanisms and the need for more research focusing on encryption as the major motivation for the attack. We hope this analysis serves as a catalyst for further advancements and a guidepost for future endeavors in combating cryptographic ransomware.

### 4.1. Lessons learned

In this paper, we reviewed cryptographic ransomware from the perspective of what we believe is its differentiating factor from other families of cyberattacks, namely encryption. Here we summarize the lessons learned in this survey. Foremost, crypto-ransomware is an increasing threat that cripples critical capabilities for both public and commercial services for an extended period of time. When we add the amounts of paid ransoms, losses are in the tens of millions of dollars. They could result in loss of human life, taking into consideration ransomware groups' "taste" for medical facilities during the time of the COVID-19 pandemic, or even involve themselves in cyber warfare. Secondly, researchers have seen a discrepancy in describing and categorizing ransomware ranging from plain malware to sophisticated cyber kill chains representing the activity of sophisticated APT-like threat actors. We observed ransomware through our cyber kill chain. Thirdly, from a research perspective, we have seen many proposals dealing with different aspects of crypto-ransomware, and most of them take a hybrid approach to deliver the solution. Fourthly, we have seen that despite the academic community's mature and focused research, commercial solutions mainly apply machine learning and anomaly detection solutions. Finally, we have seen that specialized research around the problem of encryption detection and general control of encryption operations is in the apparent minority among the research topics into crypto-ransomware.

**Table 3**

Pros, Cons and Effectiveness of different methodologies for detecting ransomware's pre-encryption and encryption activities.

| Methodology | Pros | Cons | Effectiveness |
|---|---|---|---|
| API and system call monitoring-based detection | It provides real-time detection, as system calls and APIs are used during the ransomware's activities.<br>It can reveal information about the ransomware's inner workings, which may aid in further developing or fine-tuning the countermeasures.<br>It can be combined with other detection techniques for increased accuracy.<br>Enables the identification of specific attack vectors and potentially vulnerable system components. | False positives may arise due to benign software with similar system call patterns.<br>Ransomware can evolve to avoid detection by using different APIs or obfuscating system calls.<br>It can be resource-intensive, requiring monitoring and analyzing many API and system calls.<br>It can be bypassed by advanced ransomware that utilizes unconventional techniques or exploits vulnerabilities in system calls.<br>Usage in signature-based solutions might be ineffective, as new ransomware strains may use different API and system call patterns. | This approach can be highly effective because ransomware often uses specific system calls and APIs to access and manipulate files.<br>Monitoring for unusual patterns in these calls can help identify ransomware activities. |
| I/O monitoring-based detection | It can be used as an early warning system, as I/O spikes might indicate ongoing ransomware activity.<br>It can be less resource-intensive compared to monitoring system calls and APIs.<br>It can detect ransomware even if it uses unconventional or previously unseen system calls and APIs, if the I/O patterns are consistent with encryption activities.<br>Relatively easier to implement compared to monitoring system calls and APIs.<br>Allows for the identification of affected files and systems, enabling targeted response and recovery efforts. | It may produce false positives, as high I/O activity can be generated by legitimate applications.<br>May not detect ransomware with low I/O activity, such as those that encrypt files selectively or over an extended period.<br>It can be circumvented by ransomware that employs techniques to blend its I/O activity with normal system behavior.<br>Continuous monitoring and fine-tuning of detection thresholds are required to maintain accuracy and reduce false positives. | This approach is moderately effective, as ransomware typically generates high I/O activity while encrypting large numbers of files. |
| Filesystems monitoring-based detection | It allows ransomware detection based on its unique file manipulation behavior.<br>It can provide insights into the ransomware's encryption strategy, aiding in decryption and recovery efforts.<br>It can detect ransomware that employs file-level encryption, which is a common feature in many ransomware strains.<br>It can help identify the specific encryption algorithms used by ransomware, which may aid in decryption efforts.<br>It provides opportunities for early intervention, as filesystem-related activities typically occur before actual encryption. | It may produce false positives due to benign applications with similar filesystem-related activities.<br>Ransomware can adapt its filesystem activities to bypass detection mechanisms.<br>It may not be effective against ransomware that operates at the disk or partition level, as such strains may not exhibit the same filesystem-related behavior.<br>It can be bypassed by ransomware that employs file-less techniques or unconventional filesystem access methods.<br>Requires continuous updates to the detection rules and heuristics, as new ransomware strains may exhibit different filesystem-related behavior. | This approach can be highly effective, as ransomware often exhibits specific file access and modification patterns, such as renaming files, changing file extensions, or modifying file attributes. |

Overall, there is no one-size-fits-all answer to the most effective approach. The effectiveness of each method depends on the specific ransomware and its behavior. Combining all approaches and machine learning may offer the best defense against ransomware attacks. Adding to that behavioral analysis and user awareness, effectiveness increases significantly. The information in Table 3 summarizes the pros and cons of each methodology, providing the argument for the hybrid approach advantage. These conclusions result from a review and study of different proposals presented in this survey and our research into commercial solutions that offer capabilities that could be used to detect pre-encryption and encryption activities by ransomware attacks.

### 4.2. Main findings

We focused on the Encryption phase described in our cyber kill chain and divided various methodologies into three major groups:

- API and system call monitoring-based detection
- I/O monitoring-based detection
- file system monitoring-based detection

Reviewing the work of researchers through the prism of those three detection methodologies, it was not surprising that, in the case of complete proposals for detection, researchers preferred a hybrid approach utilizing primarily combinations of the three or fewer methodologies, with machine learning being the preferred tool in many proposals.

Most of the research was focused on Windows operating systems, but Android mobile operating systems and a variety of Internet of Things applications were also present in the reviewed work.

The overall conclusion of this survey is that more methods and techniques described in the surveyed research efforts should be utilized in real-life products by trying to remove some of the common obstacles already described. Detecting pre-encryption and encryption activities provides a high level of confidence that ransomware can be intercepted before doing severe damage. Introducing some benchmark methods in the detection of encryption along with appropriate datasets is a desperate need in this area of research.

**Declaration of Competing Interest**

The authors declare that they have no known competing financial interests or personal relationships that could have appeared to influence the work reported in this paper.

**CRediT authorship contribution statement**

**Kenan Begovic:** Conceptualization, Methodology, Formal analysis, Resources, Writing – original draft, Writing – review & editing. **Abdulaziz Al-Ali:** Conceptualization, Validation, Writing – review & editing, Supervision. **Qutaibah Malluhi:** Conceptualization, Validation, Writing – review & editing, Supervision.

**Data availability**

No data was used for the research described in the article.

**Kenan Begovic** is currently a Ph.D. candidate in Computer Science at Qatar University. He received his MS in Information and Computer Security from University of Liverpool. Kenan is an awarded information security professional with over 20 years of experience in the field across the various industry verticals. He specializes in "greenfield" implementations of Information Security Management Systems. As one of the pioneers of information security in South-East Europe and the GCC region, Kenan managed information security governance, risk, and operations for intelligence/law enforcement, government departments, Islamic finance industry leaders, and sports and entertainment companies. Currently, Kenan is holding the position of Chief Information Security Officer for beIN Media Group in Doha where he man-ages information security in all of the Group's members spread across the five continents and 11 countries.

**Abdulaziz Al-Ali** received the Ph.D. degree in machine learning from the University of Miami, FL, USA, in 2016. He is currently an Assistant Professor in the Computer Science and Engineering Department, and director of the KINDI Center for Computing Research at Qatar University. In addition to developing novel Machine Learning techniques, his current research involves building predictive models for industry applications that involve textual, visual, and biometric-based data. Dr. Al-Ali was awarded several research grants from the Qatar National Research Fund as well as from the industry. He has served as an editor for data analytics related books and has published in a number of reputable peer-reviewed journals and conferences.

**Qutaibah Malluhi** is a Professor at the Department of Computer Science and Engineering at Qatar University (QU). He was the Head of the Department between 2006 and 2012 and the director of the KINDI Center for Computing Research at QU between 2012 and 2016. He served as a professor at Jackson State University. Dr. Malluhi was the co-founder and CTO of Data Reliability Inc. He was a consultant for several telecommunication companies where he built networks, designed distributed applications and developed telecommunication management software. Prof. Malluhi has received the QU research award, the JSU Technology Transfer award, the JSU faculty Excellence award, and several best paper awards. He has received MS and PhD degrees in computer science from the University of Louisiana, Lafayette, and BS and MS degrees in Computer Engineering from King Fahd University of Petroleum and Minerals, Saudi Arabia.